\documentstyle[aps,prb,multicol]{revtex}

\begin{document}
\draft
\hoffset= -2.5mm

\title{Nonequilibrium Noise in Metals at Mesoscopic Scales\cite{where}}

\author{F. Green}

\address{GaAs IC Prototyping Facility,
CSIRO Telecommunications and Industrial Physics,\\
P.O Box 76, Epping ~NSW 2121, Australia}

\author{M. P. Das}

\address{Department of Theoretical Physics,
Institute for Advanced Studies,
The Australian National University,\\
Canberra ~ACT 0200, Australia}

\maketitle

\begin{abstract}
We review a semiclassical theory of high-field current noise in
strongly degenerate conductors, based on
propagator solutions to the Boltzmann transport equation for
current fluctuations. The theory provides a
microscopic description of correlation-induced suppression
of noise in high-mobility heterojunction devices,
and is also applicable to diffusive mesoscopic
structures. We discuss the behaviour of thermal noise
in a mesoscopic wire, from equilibrium to the high-field limit..
\end{abstract}

\pacs{Keywords:
{\em nonequilibrium fluctuations, noise, 
electron correlations, mesoscopics}
}

\begin{multicols}{2}

Large departures from equilibrium are a major
issue in transport theory,\cite{wrf,jh} more so in the
mesoscopic realm since, at nanometre scales,
very modest external voltages easily generate intense
driving fields with highly nonlinear effects.\cite{bw}
The problem is of greater severity for many-body
properties such as current noise,\cite{swI,swII}
and in this paper we study aspects of high-field
current fluctuations for degenerate electrons,
an area scarcely explored.
We outline a microscopic approach
to fluctuations in metallic systems
far from equilibrium, illustrating this
with an application to real heterojunction devices
\cite{gc}
and finally with an analysis of high-field thermal noise
in an idealised, diffusive mesoscopic wire.
\cite{l+djb}

\section{Formalism}

\subsection{Transport}

Our method generalises the purely classical
Boltzmann-Green-function theory
of Stanton and Wilkins.\cite{swI,swII,swt}
We start with the one-body transport equation.
For simplicity consider a uniform conductor.
The semiclassical one-electron Boltzmann equation is

\begin{eqnarray}
{\left[
{\partial\over {\partial t}}
- {{e{\bf E}}\over {\hbar}}
{\bbox \cdot}
{\partial\over {\partial {\bf k}}}
\right]} f_{s,{\bf k}}(t)
=&& {\cal C}_{s,{\bf k}}[f(t)],
\label{eq1}
\end{eqnarray}

\noindent
where the time-dependent distribution $f_{s,{\bf k}}(t)$
is labelled by band-and-spin index $s$ and wave vector
${\bf k}$; the local driving field (uniform here)
is ${\bf E}$. The collision operator ${\cal C}_{s,{\bf k}}[f]$
carries, at the semiclassical level, the effects of microscopic
scattering from all sources.
In a degenerate system, ${\cal C}$ is inherently nonlinear in $f$.
Normalisation to the density $n$ is such that
$n = \Sigma_{s,{\bf k}} f_{s,{\bf k}}/\Omega $
in a system of volume $\Omega$.

In the steady state, Eq. (\ref{eq1}) maps
the equilibrium Fermi-Dirac distribution $f^{\rm eq}$
to $f$. Introducing the difference function
$g \equiv f - f^{\rm eq}$, we have

\begin{eqnarray}
- {{e{\bf E}}\over {\hbar}}
{\bbox \cdot}
{{\partial g_{s,{\bf k}}}\over {\partial {\bf k}}}
=&& 
{{e{\bf E}}\over {\hbar}}
{\bbox \cdot}
{{\partial f^{\rm eq}_{s,{\bf k}}}\over {\partial {\bf k}}}
+ {\cal C}_{s,{\bf k}}[f^{\rm eq} + g].
\label{eq2}
\end{eqnarray}

\subsection{Steady-State Fluctuations}

The equilibrium density-density fluctuation for free electrons is
$\Delta f^{\rm eq} \equiv k_BT \partial f^{\rm eq}/\partial \varepsilon_F$,
where $k_BT$ is the thermal energy and $\varepsilon_F$
is the chemical potential.
Eq. (\ref{eq2}), linearised, maps $\Delta f^{\rm eq}$
adiabatically to its nonequilibrium counterpart. Solving
for the fluctuation difference $\Delta g$ according to

\begin{eqnarray}
- {{e{\bf E}}\over {\hbar}}
{\bbox \cdot}
{{\partial \Delta g_{s,{\bf k}}}\over {\partial {\bf k}}}
=&& 
{{e{\bf E}}\over {\hbar}}
{\bbox \cdot}
{{\partial \Delta f^{\rm eq}_{s,{\bf k}}}\over {\partial {\bf k}}}
\cr
{\left. \right.} \cr
&& + 
\sum_{s',{\bf k'}}
{ {\delta {\cal C}_{s,{\bf k}}}\over {\delta f_{s',{\bf k'}}} }
{\Bigl(
\Delta f^{\rm eq}_{s',{\bf k'}} + \Delta g_{s',{\bf k'}}
\Bigr)},
\label{eq3}
\end{eqnarray}

\noindent
the complete steady-state fluctuation
can be constructed as $\Delta f \equiv \Delta f^{\rm eq} + \Delta g$.
The effect of degeneracy on $\Delta g$ is explicit, 
since $\Delta f^{\rm eq} = f^{\rm eq}(1 - f^{\rm eq})$.
Moreover, the leading right-hand term of Eq. (\ref{eq2})
for $g$ itself contains
{~$\partial f^{\rm eq}/\partial {\bf k} \propto \Delta f^{\rm eq}$}. Thus
{\em the very nature of the equilibrium state imposes
a connection between the nonequilibrium one- and two-particle
structures}, $g$ {\em and} $\Delta f$.
\cite{fg}

\subsection{Dynamic Fluctuations}

These are obtained from the linearised form of Eq. (\ref{eq1}) by
calculating its retarded resolvent $R(t)$,
\cite{ks}
with which one then constructs
the transient current autocorrelation $C_{xx}(t)$:

\begin{eqnarray}
R_{s s'; {\bf k} {\bf k'}}(t - t') \equiv&&
\theta(t - t')
{ {\delta f_{s,{\bf k}}(t)}\over {\delta f_{s',{\bf k'}}(t')} }
\label{eq4}
\end{eqnarray}

\noindent
and

\begin{eqnarray}
C_{xx}(t) \equiv&& {e^2\over l^2}
\sum_{s,{\bf k}} \sum_{s',{\bf k'}}
(v_x)_{s,{\bf k}}
{\Bigl(
R_{s s'; {\bf k} {\bf k'}}(t)
\Bigr.} \cr
&& -
{\Bigl.
R_{s s'; {\bf k} {\bf k'}}(\infty)
\Bigr)}
(v_x)_{s',{\bf k'}} \Delta f_{s',{\bf k'}},
\label{eq5}
\end{eqnarray}

\noindent
where $(v_x)_{s,{\bf k}}$ is the group velocity
in the direction of $-{\bf E}$,
and $l$ is the sample length.  Eq. (\ref{eq5}) extends
the classical definition of Stanton and Wilkins\cite{swI}
to degenerate systems, with the following interpretation.
At $t = 0$ a spontaneous fluctuation $v'_x \Delta f'$
perturbs the flux in its steady state. Its subsequent fate is
determined by the propagator $R(t)$.
After removal of the steady-state asymptote at $t \to \infty$,
the product
$(-ev_x/l) R (-ev'_x/l) \Delta f'$
gives the dynamical
current-current correlation.

Our premise is that fluctuations in the electron gas
are induced almost instantaneously compared with their
relaxation rate. This allows a two-step calculation.
First, compute the strength of the spontaneous
fluctuations in the steady state
(with the underlying equilibrium statistics included manifestly).
Second, derive their semiclassical dynamics
and hence the noise spectrum.
This prescription satisfies the fluctuation-dissipation theorem.
\cite{fg}

\section{Consequences}

The simplest form of our theory is based on the Drude model
in a single parabolic conduction band, for which
${\cal C} \equiv - g/\tau$; at room temperature
the inelastic collision time $\tau$ is of order $10^{-13}{\rm s}$.
The noise spectral density is
\cite{swI,gc}

\begin{eqnarray}
S(\omega)
\equiv&& 4 \int^{\infty}_0\!\! C_{xx}(t) \cos(\omega t) dt
\cr
{\left. \right.} \cr
=&& {{4Gk_BT}\over {1 + \omega^2 \tau^2}}
{\left[
1 + {\left( {{\Delta n}\over n} \right)}
{{m^* \mu^2 E^2}\over {k_BT}}
\right]},
\label{eq6}
\end{eqnarray}

\noindent
in which $G$ is the sample conductance,
$m^*$ the effective mass, $\mu = e\tau/m^*$ the mobility, and
$\Delta n = 2\Sigma_{\bf k} \Delta f_{\bf k}/\Omega$.
For $E = 0$ in the static limit, $S(0)$ is
the Johnson-Nyquist noise. At high fields
the ``hot-electron'' term $\propto E^2$
in Eq. (\ref{eq6}) is quite strongly suppressed by degeneracy;
while a classical system has $\Delta n/n = 1$,
a $\nu$-dimensional degenerate system
has, instead, $\Delta n/n = \nu k_BT/2\varepsilon_F \ll 1$.

Equations (\ref{eq3}), (\ref{eq5}), and (\ref{eq6})
show $S$ to be a linear functional of $\Delta f^{\rm eq}$. Insofar as
Coulomb and exchange interactions modify the 
the free-particle form of the equilibrium
density-density fluctuation, it follows that
{\em thermal noise carries a signature of
the internal correlations of the electron gas}.
Apart from its physical significance,
this has immediate practical implications.

\subsection{Device Noise}

An important consequence for microwave technology
is reduction of thermal noise in a two-dimensional (2D)
electron gas,\cite{gc}
confined at the heterojunction of a
high-electron-mobility transistor (HEMT).
\cite{wv}
Basically, the occupancy within a HEMT is
$f^{\rm eq}_{\bf k} =
\{ 1 + \exp[(\varepsilon_{\bf k} + \varepsilon_0(n)
- \varepsilon_F)/k_BT] \}^{-1}$,
where $\varepsilon_{\bf k}$ is the 2D band energy and $\varepsilon_0(n)$
is the ground-state energy in the heterojunction quantum well.
The density governs $\varepsilon_0$ because
the mean-field electronic potential in the well
is self-consistent. As a result $\Delta f^{\rm eq}$ is renormalised
to $\gamma {\Delta f}^{\rm eq}$ by the suppression factor
\cite{gc}
$\gamma(n) = [1 + (d\varepsilon_0/dn)\Delta n/k_BT]^{-1}$.

The observable noise is therefore reduced by
as much as 65\% for sheet densities
of $10^{12} {\rm cm}^{-2}$, typical in conductive channels.
Under normal conditions this means that the effective noise
temperature in a HEMT is
{\em 100K}, not the 300K of a bulk conductor.
Such suppression is unique to self-consistently quantised systems.
Circuit-theoretical arguments show that, in production
devices, it should lead to
an extrinsic noise figure which is
half that of bulk field-effect transistors,
of otherwise similar circuit performance.\cite{gc}
This agrees well with real device comparisons.\cite{wrb}

\subsection{Mesoscopic Noise}

Sample lengths approaching the mean free path $\lambda$
take us into the mesoscopic regime.
\cite{l+djb}
Even for an embedded
slice of long, spatially uniform conductor,
the propagator $R(t)$ acquires spatial structure
below $\lambda$, satisfying

\begin{eqnarray}
&&{\left[
{\partial\over {\partial t}}
+ {\bf v}_{s,{\bf k}}
{\bbox \cdot}
{\partial\over {\partial {\bf r}}}
- {{e{\bf E}}\over {\hbar}}
{\bbox \cdot}
{\partial\over {\partial {\bf k}}}
\right]}
R_{s s'; {\bf k} {\bf k'}}({\bf r} - {\bf r'}, t - t')
\cr
{\left. \right.} \cr
&& {~} =
\Omega \delta_{{\bf k} {\bf k'}} \delta_{s s'}
\delta({\bf r} - {\bf r'}) \delta(t - t')
\cr
{\left. \right.} \cr
&& {~~~} + \sum_{s'',{\bf k''}}
{ {\delta {\cal C}_{s,{\bf k}}}\over {\delta f_{s'',{\bf k''}}} }
R_{s'' s'; {\bf k''} {\bf k'}}({\bf r} - {\bf r'}, t - t').
\label{eq7}
\end{eqnarray}

\noindent
The sums in Eq. (\ref{eq5}) for $C_{xx}$ now include
integrals over the mesoscopic slice, while
the macroscopically homogeneous propagator of Eq. (\ref{eq4})
re-emerges as $\int\!d^\nu r R/\Omega$.

Using Eqs. (\ref{eq7}), (\ref{eq5}), and (\ref{eq6})
we have computed the thermal-noise spectrum
for an embedded uniform 1D wire, in the Drude model
previously studied by Stanton\cite{swt} within the
classical limit $k_BT \gg \varepsilon_F$.
The example, although suggestive, is artificial not least because
``wire'' and ``leads'' are operationally indistinguishable.

We examine the degenerate limit, for which $\lambda = \tau v_F$
in terms of the Fermi velocity. At zero frequency the
equilibrium noise is given by

\begin{equation}
S^{\rm eq}(0) = 4Gk_BT
{\left[ 
1 - {\lambda\over l}(1 - e^{-l/\lambda})
\right]},
\label{eq8}
\end{equation}

\noindent
where $l$ is the length of the wire segment, much smaller
than the enclosing 1D ``volume'' $\Omega$. In the limit $l \ll \lambda$
Eq. (\ref{eq8}) reduces to $S^{\rm eq}(0) = (4e^2/\pi \hbar)k_BT$.
Since this is the ballistic regime we note that the thermal noise
scales with $e^2/\hbar$,
the universal unit of conductance.\cite{l+djb} 

On the other hand the nonequilibrium ballistic noise is

\begin{equation}
S(0) = {{4e^2}\over {\pi \hbar}}k_BT
{\left[
1 + {\left( 1 + {{\mu E}\over v_F} \right)}
            \exp{\left( -{{2v_F}\over {\mu E}} \right)}
\right]}.
\label{eq9}
\end{equation}

\noindent
The expression is nonperturbative in $E$
owing to nonanalyticity of the solutions to
Eqs. (\ref{eq2}) and (\ref{eq3})
for a uniform nonequilibrium system.
\cite{bw,swt}
Our exact, if simple, model calculation shows
that one cannot take for granted
the existence of an expansion for the fluctuations
near the equilibrium state, in powers of the field.

At large fields the ballistic noise becomes linear:
\cite{swt}

\begin{eqnarray}
S(0) =&& {{4e^2}\over {\pi \hbar}}k_BT
{\left( {{\mu E}\over v_F} \right)} + {\cal O}(E^{-2})
\cr
{\left. \right.} \cr
\rightarrow&& 4Gk_BT{\left( {{eV}\over {4\varepsilon_F}} \right)},
\label{eq10}
\end{eqnarray}

\noindent
where $V = El$ is the voltage across the wire
(still with $l \ll \lambda$),
and we have reinstated the Johnson-Nyquist normalisation.
Two remarks can be made on this equation:

$(a)$ the linear dependence on $E$ is kinematic, reflecting the
shift of the centroid of the fluctuation distribution
in $k$-space by $eE\tau/\hbar$. In contrast,
the quadratic dependence of the macroscopic noise in
Eq. (\ref{eq6}) is dissipative, and sensitive to
thermal broadening of $\Delta f_k$ over the bulk.

$(b)$ The second form of Eq. (\ref{eq10})
is equivalent to $S(0) = 2eI(\Delta n/n)$,
where $I = GV$ is the current through the wire.
Thus, ballistic thermal noise in the 1D Drude approximation
is given by the classical shot-noise formula
\cite{l+djb,swt}
attenuated by the degeneracy, a result valid for all
densities and temperatures.

While thermal noise must go to zero with the temperature,
this is not the case for true shot noise, whose non-thermal
origin is the random transit of individual carriers
across the sample.
It is beyond our present scope to discuss applications of the
Boltzmann-Green-function formalism to diffusive mesoscopic shot noise,
\cite{l+djb,smd,sbkpr}
which we are actively investigating.

\section{Summary}

We have outlined a semiclassical framework for calculating thermal
fluctuations in metallic electron systems far from equilibrium.
Our approach also describes how Coulomb and exchange correlations,
present at equilibrium, appear in the
nonequilibrium current noise. The consequences
for device design are exemplified by the physics of
correlation-induced noise suppression
in heterojunction field-effect transistors.

The same formalism can be applied to
diffusive mesoscopic noise. An illustrative 1D model provides
evidence that mesoscopic noise may not always have a perturbation
expansion at low fields, if the underlying distributions
are nonanalytic in the equilibrium limit.
At high fields the model recovers the shot-noise-like behaviour of
ballistic thermal noise.\cite{swt}
For a metallic wire we find that this is
attenuated in proportion to the degeneracy of the system.

\end{multicols}

\end{document}